\documentclass{icrc29}
\usepackage{graphicx,amssymb,amsmath,times}
\begin{document}
\title[Using ACT arrays as Intensity Interferometers]{Using Atmospheric
  Cherenkov Telescope arrays \\ as Intensity Interferometers}
\author[S.LeBohec and J.Holder] {S. LeBohec$^a$ and J. Holder$^b$\\
        (a) Physics Department, University of Utah, 115 S. 1400 E. Suite 201, 
            Salt Lake City, Utah 84112-0830, USA\\ 
        (b) School of Physics and Astronomy, University of Leeds, UK
        }
\presenter{Presenter: S. LeBohec (lebohec@physics.utah.edu)}

\maketitle

\begin{abstract}
The Narrabri intensity interferometer was successfully used until 1974 to
observe 32 stars \cite{brown1974}, all brighter than B=+2.5, among which some
were found to have an angular diameter as small as 0.41+/-0.03
milli-arc-seconds (mas). The technique was then abandoned in favor of
Michelson interferometry \cite{michelson, michelsonpease}. Here we consider
the technical feasibility and scientific potential of implementing intensity
interferometry on Imaging Air Cherenkov Telescope arrays. The scientific 
motivations are varied, including stellar diameter measurements and 
investigations of the circumstellar environment. Long baselines and short 
wavelengths are easily accesible to this technique, making it uniquely suited 
for some applications.

\end{abstract}
\vspace{-0.4cm}
\section{Intensity Interferometry}
In a stellar intensity interferometer\cite{rhbbook}, the light from a star is
received by two separated photoelectric detectors through a narrow-band 
filter. The technique relies on the fact that the current output fluctuations 
$\Delta i_1$ and $\Delta i_2$ of the two detectors are partially correlated. 
The principal component of the fluctuation is the classical shot noise which 
does not show any correlation between the two detectors. In addition, there is 
a smaller component, the wave noise, that shows some correlation. Because of
this correlation, the product of the fluctuations in each detector
will be positive, and provide a measurement of the square of the degree 
of coherence $\rm \gamma_d$ of the light at the two detectors.
$\rm \gamma_d$ is equivalent to the fringe visibility measured with a
Michelson interferometer. It depends upon the angular diameter $\rm \theta$ 
of the star, the wavelength $\rm \lambda$ and the distance between the two 
detectors $\rm d$ . In the case of a star modeled as a uniform disk it
reaches zero when $\rm \rm d=1.22\lambda/\theta$. In fact, according to the
van Citter-Zernike theorem, $\rm \gamma_d$, the complex degree of
coherence is the normalized Fourier transform of the source intensity
distribution projected on a line parallel to the line joining the two
detectors. Measuring $\rm |\gamma_d|^2$ provides information on the angular
structure of the source.    

The shot noise from both channels is responsible for most of the correlator
output fluctuations to which the wave noise correlation has to be compared
for the sensitivity of a specific experiment to be estimated. The signal to 
noise ratio can be expressed as 
$(S/N)_{RMS}=A\alpha n |\gamma_d|^2 (\Delta fT/2)^{1/2}$
where A is the collection area of each telescope, $\alpha$ the photo-detector 
quantum efficiency, $\rm \Delta f$ the bandwidth of the electronics including
the photo-detector, T the integration time, n the intensity of the source in
photons per unit optical bandwidth, per unit area and per unit time.  Using 
$\rm 100~m^2$ telescopes with 30\% quantum efficiency photo-detectors, 1~GHz 
electronics during a full 10 hours night would permit to measure the 
diameter of stars with magnitude less than 8. 
\vspace{-0.3cm}

\subsection{ Comparison with Michelson Interferometry}
After decades of development, a number of world-class instruments for
long-baseline optical interferometry are currently operating or soon to come
online (VLTI, Keck interferometer). These are, without exception, based upon
the Michelson technique for interferometry. The advantages of this method are
clear. As it relies on the visibility of fringes produced by the amplitude 
interference formed by the light collected by two telescopes, it permits 
measurements of stars much dimmer than intensity interferometry with same size
telescopes. Furthermore, with intensity interferometry, only the modulus of 
the coherence is measured and the two-dimensional image can only be 
reconstructed up to a central symmetry.

Unfortunately, Michelson interferometry is also a very challenging technique
as the relative lengths of the light paths have to be controlled to an
accuracy smaller than the wave length of the light being measured. This
requires high optical quality and high precision tracking. The situation is
further complicated by the effects of atmospheric turbulence which must be
actively compensated for. These difficulties have constrained Michelson
interferometry to small baselines (most interferometers provide baselines of
less than 100m) and long wavelengths (most interferometers work at more than
1$\mu$m) while the maximal angular resolution is proportional to the base line
and to the inverse of the wavelength.
 
Intensity interferometry, on the contrary, only requires control of the light
paths to an accuracy fixed by the light coherence time ($\sim\rm100ns$). It is
therefore insensitive to atmospheric fluctuations and easy to implement at
shorter wavelengths. In addition, as the correlation is made \textit{after}
the photons are detected, intensity interferometry also permits
simultaneous measurement of $|\gamma_d|^2$ between any two-telescope pair
of an array, while with Michelson interferometry this is impossible without a
loss in sensitivity. The major drawback associated with intensity
interferometry is the need for very large quantities of light.  The
necessary large light collectors do not however need to be of optical
astronomy quality as they are only required to isochronally concentrate the
light on a photo-detector.
\vspace{-0.4cm}

\section{Possible Implementations}

The Narrabri interferometer consisted of two telescopes 6.5m in diameter with
an 11m focal length. The telescopes were carried on trucks running on a
circular railway track 188m in diameter. This allowed for a baseline from 10m
to 188m to track any star while keeping the line joining the two telescopes
perpendicular to the direction of the star in such a way that no delay was
required to bring the signals in time. At the focus of each telescope, the
converging light was collimated and passed through an interference filter
centered on 443nm with a passing bandwidth of 10nm. The measured light
intensity was converted to a currant by a photomultiplier tube of 25\% quantum
efficiency at 440nm and 60MHz effective bandwidth. The signals were sent to
the control building where the correlator, a transistor based linear
multiplier, was located.

While the Narrabri instrument was successful, significant sensitivity
improvements can be achieved by observing the same star with a full array of
telescopes providing measurements over several baselines simultaneously
\cite{herrero}. The number of baselines is proportional to the square of the
number of telescopes in the array. This could be combined with the
technological developments during the last 30 years which provide higher
quantum efficiency photo-detectors, higher bandwidth photo-detectors and
electronics and the possibility of processing digitized signals at high speed.

Modern ground-based $\gamma$-ray observatories (CANGAROO,
HESS, MAGIC and VERITAS)
consist of arrays of IACTs which satisfy many of the specifications for an
intensity interferometer.  The characteristics of several IACT arrays are
summarized in Table 1. We have included HESS-16 in order to illustrate the
benefits of a large array. IACT arrays typically extend over $\sim\rm200m$
making them comparable to the Narrabri interferometer in terms of the angular
resolution they could achieve. The telescopes have diameters ranging from 10~m
to 17~m providing a gain of a factor of 2.8 to 8.0 in sensitivity (1 to 2.3
magnitudes) when compared to the Narrabri interferometer. Furthermore, these
arrays, with one exception, are made up of 4 telescopes and so permit
measurements over up to 6 baselines simultaneously (120 in the case of
HESS-16). For the measurement of a symmetric object this corresponds to a
sensitivity gain of $\sim2.5$ or one magnitude ( $\sim10$ or 2.5 magnitudes
for HESS-16). Figure~\ref{stars} shows $\rm |\gamma_d|^2$ as a function of
baseline for stars of different angular diameters. Measurements at only two
points along this curve enable $\theta$ to be measured unambiguously.

\begin{table}
\begin{center}
\caption{Characteristic of the major IACT arrays compared to the Narrabri interferometer. 
$\rm N$ is the number of telescopes, $\rm A$ is the collection area of each telescope, 
$\rm \Delta t$ is the dime dispersion introduced by the optics, $\rm n \times d$ indicate 
the number and the length of available baselines for n observation at zenith, 
$\rm \theta_{Min}$ is the corresponding angular resolution ($\rm \gamma_d=0.5$) for 
observations at 400~nm and $V_{Max}$ is the magnitude of the faintest non resolved star 
($\rm \gamma_d=1$) that can be measured in one night using all available baselines. 
\label{chertel}}
\begin{tabular}{||c|c|c|c|c||}\hline 

Array    & N & A ($\rm m^2$)&  Baselines  \\\hline \hline

MAGIC    & 2 & 227   & $\rm 1\times \sim85m$  \\\hline
CANGAROO & 4 &  78   & $\rm 5\times \sim100m$ \\
         &   &       & $\rm 1\times 184m$     \\\hline
VERITAS  & 4 & 113   & $\rm 3\times 80m$      \\
         &   &       & $\rm 3\times 140m$     \\\hline
HESS     & 4 & 113   & $\rm 4\times 120m$     \\
         &   &       & $\rm 2\times 170m$     \\\hline
HESS16   &16 & 113   & 120 $\rm \times~from~120m$ to  $\rm 510m $ \\\hline
Narrabri & 2 & $\rm 30m^2$  & $\rm 1 \times 10m \to 188m$ \\\hline
\end{tabular}
\end{center}
\end{table}
IACTs can be used for
gamma-ray observations only during Moonless nights. Stray light from the Moon
will only marginally reduce the sensitivity of an intensity interferometer and
telescope time allocation could be set according to the Moon visibility,
leaving almost half the night time available for interferometry measurements.

Conversion of an IACT into an intensity interferometer receiver would require
the installation of a narrow-band filter in front of only one of the
photo-multipliers of the camera. In one possible implementation, signals from
this channel could be transferred to a central location via an analog optical
fiber. After being digitized at high rate, the signals could be duplicated
and aligned in arrival time for each pair in the array before being multiplied
and integrated over time. Field programmable gate arrays seem to be the ideal
tool for implementing such a system.

Existing IACT arrays have the inconvenience of not having at least one 
pair of telescopes close together to allow the measurement of the degree 
of coherence from a non resolved source. This could be compensated for 
by splitting each optical channel in two in such a way the degree of 
coherence at each telescope can be measured. Another important difference 
between the Narrabri interferometer and IACT arrays is the fact that in the 
latter, telescopes are at fixed locations while at Narrabri, the two 
telescopes could be moved along tracks to keep the signals aligned in time and 
to maintain the fixed baseline as a star was tracked for long periods of 
time. While with fixed telescopes the baseline will unavoidably change during
an observation, analog or even digital programmable delays can be used to
align the signals in time and the varying baseline can be used to our
advantage, providing additional coverage of the phase space. 
\vspace{-0.4cm}   
\section{Science}
\begin{figure}[stars]
\begin{center}
      \includegraphics*[height=7cm]{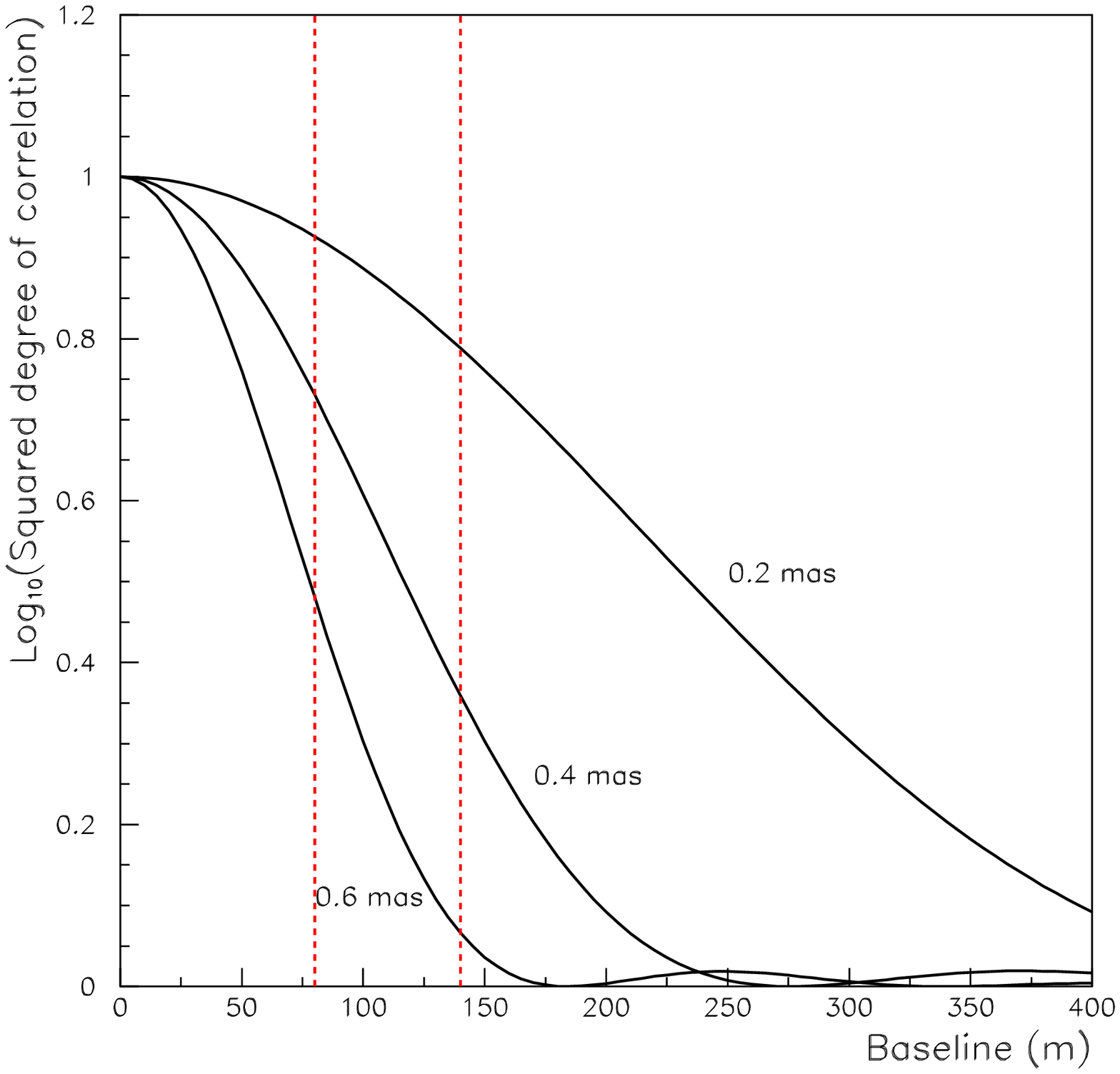}
      \hspace{0.5cm}
      \includegraphics*[height=7cm]{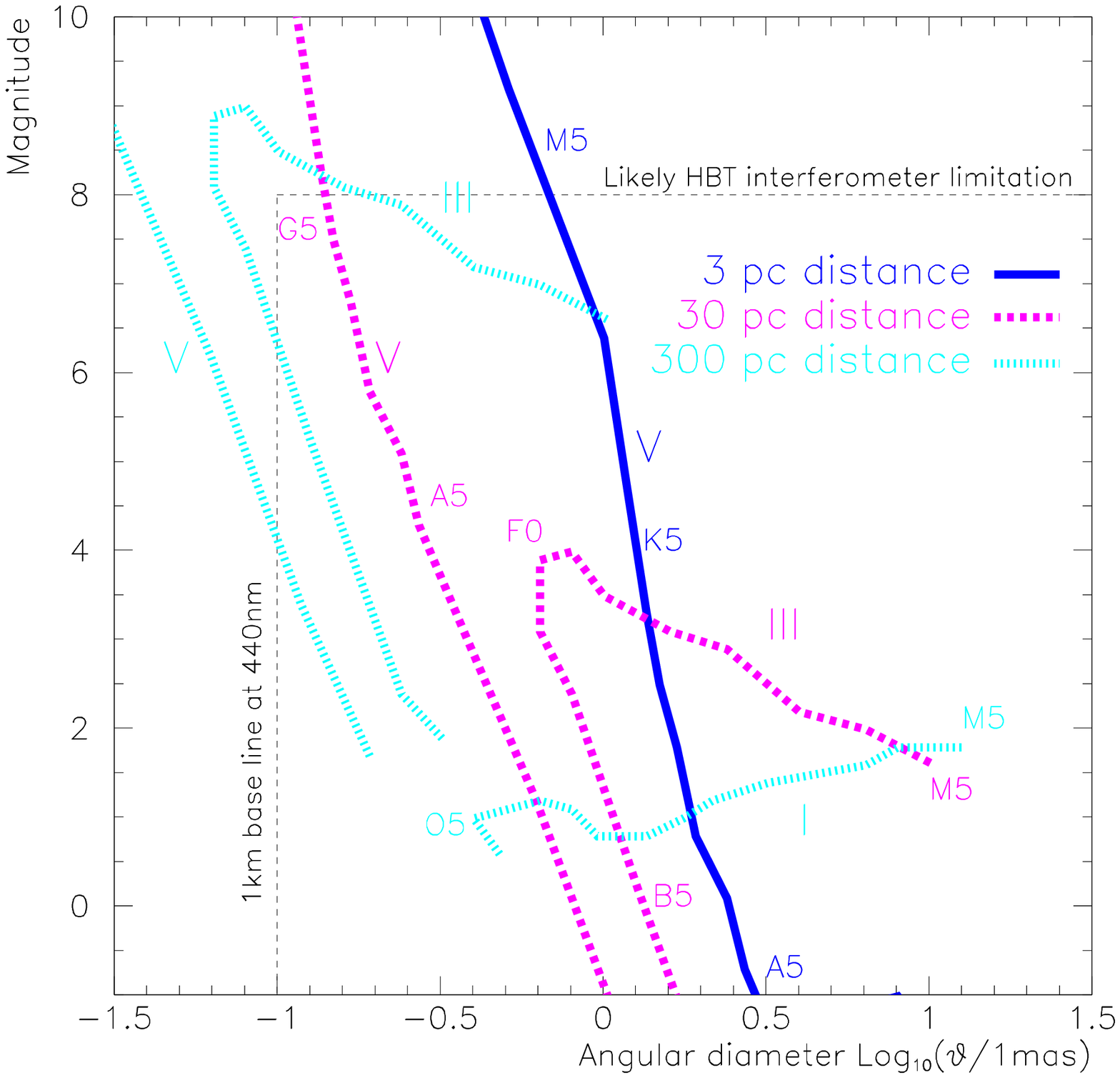}
\caption{\label {stars} {\bf Left:} $\rm |\gamma_d|^2$ as a function of
  baseline for three different stellar angular diameters. The vertical lines
  indicate the two baselines available to VERITAS. {\bf Right:} Visual
  magnitude angular diameter relationship for the main sequence, the giant and
  the super-giant branches for distances of 3~pc, 30~pc and 300~pc.}
\end{center}
\end{figure}

Astrophysics which can be addressed covers many of the areas currently being
explored by Michelson interferometers (e.g. \cite{monnier}). The catalog of directly measured
stellar diameters is still limited to hundreds, and only a few measurements of
have been made of lower-mass dwarf stars and hot main sequence stars. Such
direct measurements are important for models of stellar atmospheres and
stellar evolution, and can be used to calibrate surface brightness
relations. Figure~\ref{stars} shows the magnitude-diameter relationship for
stars at different distances, indicating the wide range of stellar types
within the sensitivity range of a modern intensity interferometer.

Other applications include measuring the parameters of binary systems,
measurements of the circum-stellar environment - particularly in the case of
variable stars (e.g. Cepheids or Mira variables) or stars with high mass loss
(e.g. Be stars). It may even be possible to resolve simple features on the
surface of giant stars such as Betelgeuse. The specifics of the technique make
certain applications more attractive; long baseline measurements at short
wavelengths are very difficult for Michelson instruments but relatively easy
for intensity interferometers, as are long term monitoring of sources and
studies over a wide range of wavelengths.
\vspace{-0.3cm}

\vspace{-0.3cm}
\section{Conclusion}
We discussed the feasability of Implementing a modern-day intensity 
interferometer on Imaging Air Cherenkov Telescope arrays. Developments in 
fast digital signal processing technology now make such an instrument 
relatively easy to construct, as well as improving the sensitivity.  
Measurements at short wavelength ($<\rm400nm$) with long baselines 
($\rm \sim 1000~m$) which are still very challenging for Michelson 
interferometers could be made. Such a project could operate during bright 
moon periods, providing valuable scientific output for relatively small 
expense and no impact on the gamma-ray observing schedule.

\end{document}